\documentclass[]{raa}
\newcommand{\etal}{{\it et al.}}
\usepackage{natbib}
\usepackage{color}
\usepackage{graphicx,times}

\begin{document}
\title{MHD Seismology of a loop-like filament tube by observed kink waves}

\author{V.~Pant
    \inst{1}
        \and A.K.~Srivastava
    \inst{2}
        \and D.~Banerjee
    \inst{1,3}
        \and M.~Goossens
    \inst{4}
        \and P.F.~Chen
    \inst{5,6}
        \and N.C.~Joshi
    \inst{7}
        \and Y.H.~Zhou
    \inst{5,6}}
\institute{Indian Institute Of Astrophysics, Bangalore-560 034, India\\
           \and Department of Physics, Indian Institute of Technology (BHU), Varanasi-221005, India\\
           \and Center of Excellence in Space Sciences, IISER Kolkata, India \\
           \and Centre for Mathematical Plasma Astrophysics, Department of Mathematics, KU Leuven, Celestijnenlaan 200B, B-3001 Leuven, Belgium \\
           \and School of Astronomy and Space Science, Nanjing University, Nanjing, China \\
           \and  Key Lab of Modern Astronomy and Astrophysics, Nanjing University, Nanjing, China \\
           \and School of Space Research, Kyung Hee University, Yongin, Gyeonggi-Do, 446-701, Korea\\ }

\date{Received~~XXXXXXX; accepted~~XXXXXXX}

\abstract{We report and analyze the observational evidence of global kink oscillations in a solar filament as observed in H$\alpha$ by National Solar Observatory ({\it NSO})/Global Oscillation Network Group ({\it GONG}) instrument. An M1.1-class flare in active region 11692 on 2013 March 15 induced a global kink mode in the filament lying in the south-west of AR11692. We find periods of about 61--67 minutes and damping times of 92--117 minutes at three vertical slice positions chosen in and around the filament apex. We find that the waves are damped. From the observed global kink mode period and damping time scale using the theory of resonant absorption we perform prominence seismology. We estimate a lower cut-off value for the inhomogeneity length-scale to be around 0.34--0.44 times the radius of the filament cross-section.
\keywords{Sun: oscillations --- Sun: magnetic fields --- Sun:filament}
}
\maketitle

\section{Introduction}
Solar prominences manifest various kinds of magnetohydrodynamic (MHD) oscillatory motions \citep{Arregui2012}. These oscillations are broadly classified either as large amplitude oscillations where prominence as a whole oscillates with a velocity amplitude of a few tens of km s$^{-1}$ or as small amplitude oscillations localized in a part of a prominence and with a velocity amplitude of about 2--3 km ~s$^{-1}$. Many cases of prominence thread oscillations with small amplitudes have also been reported \citep{Yi1991,YiE1991}. Horizontal flows with simultaneous transverse oscillations were reported by \citet{Lin2002}, \citet{Lin2005} and \citet{2007Sci...318.1577O}. \citet{2007Sci...318.1577O} reported the propagating Alfv\'en waves in the filament threads and calculated the wave velocity to be about 1050 km s$^{-1}$. However, later \citet{Terradas2008} correctly interpreted them as standing kink waves and calculated the lower bound of Alfv\'en velocity in these threads with a lower value of 120 km s$^{-1}$. \citet{Lin2009} also reported the propagating kink waves in the filament threads. \citet{Ning2009} found vertical and horizontal oscillations in the prominence threads with simultaneous drifting in the plane of sky.

There have been numerous reports on small amplitude oscillations as summerized in a review article by \citet{OB02} and large amplitude oscillations \citep{Tripathi09}. \citet{Ramsey66} reported the observational evidence of large amplitude oscillations induced by disturbances coming from a nearby flare. Recent observations of large amplitude oscillations were reported in H$\alpha$ \citep{Eto2002,2003ApJ...584L.103J,2004ApJ...608..1124,Jing2006,Liu2013}, in Extreme Ultraviolet \citep{Isobe2006} and in He 10830 infrared emissions \citep{Gilbert2008}. Most recent report was by \citet{Hershaw2011} in which they investigated two large amplitude damped transverse oscillations in an EUV prominence on the solar limb. The large amplitude oscillations can be either longitudinal \citep{2003ApJ...584L.103J,Jing2006,Vrsnak2007,li12,luna12a,luna12b,zhang12,luna14} or transverse \citep{Isobe2006,2004ApJ...608..1124,schm13}. Detection of magnetohydrodynamic (MHD) waves and oscillations provide important input in diagnosing the local plasma conditions of the solar corona  by applying the principle of MHD seismology \citep{2005LRSP....2....3N,andries05,andries09,ruderman09}. The fundamental mode of the global kink wave is characterized by a displacement of the magnetic flux tube where all its parts are in phase \citep{1999Sci...285..862N}. In the MHD regime, the global transverse oscillations of magnetic flux tubes are interpreted as nearly incompressible fast kink modes \citep{edwin1983,roberts84,erdelyi08,van08a,goossens09}. 

Prominence oscillations are often found to be damped \citep{Ruderman2002,erdelyi08,andries09,arre11}. Damping is observed in both large amplitude oscillations where the prominence oscillates as a whole \citep{Hershaw2011} and at smaller scales, where different parts of the prominence show different damping timescales \citep{Terradas2002,Lin2009}.

Besides the oscillation period, the damping timescale also provides fundamental information about the physical conditions around the prominence. The MHD seismology of the localized plasma environment of the magnetic flux tubes based on both the oscillation period and damping timescale of transverse waves is performed in a consistent manner by \citet{2006RSPTA.364..433G,arre07,2008IAUS..247..228G,2008A&A...484..851G,2013ApJ...777..158S}. These fast kink radial modes get attenuated likely due to the resonant absorption and are converted into dominantly torsional (azimuthal) highly incompressible motion \citep{Ruderman2002,2002ESASP.505..137G,2006RSPTA.364..433G,2008A&A...484..851G,2013ApJ...768..191G,2008IAUS..247..228G,arre08}. However, other mechanisms (e.g., dissipation through wave leakage, curvature effects, phase-mixing, etc.) can also be at work in dissipating such waves in the solar coronal tubes \citep[e.g,][]{Ruderman2002, 2002ApJ...580L..85O,2013Ap&SS.345...25S}.

In this paper, we present evidence of large amplitude, long period damped transverse oscillations in a Solar filament as observed by {\it GONG} instrument of National Solar Observatory ({\it NSO}) in H$\alpha$. We interpret them as global kink waves as they displace the prominence tube as a whole in the transverse mode. Such an analysis has been carried out by \citet{Hershaw2011}, but they have reported the kink oscillations in EUV and there were differences in the periods of the two legs of the prominence. With that, the two legs of the prominence oscillated in phase initially and then gradually moved out of phase. \citet{2004ApJ...608..1124} and \citet{Eto2002} observed winking filaments in H$\alpha$ with intermediate periods (20--40 min). However, there was no signature of damping. \citet{Isobe2006} and \citet{Chen2008} also reported the undamped transverse oscillations, which happened during or immediately prior to eruption. \citet{Gilbert2008} reported long period, large amplitude vertical oscillations but the damping time observed by them is six times the period of the oscillation.
Prominence seismology in large amplitude oscillations has been carried out by \citet{Pinter2008} and \citet{Vrsnak2007}, and in prominence threads by \citet{Lin2009}.
The paper is organized as follows: In Sect. 2, we present briefly the observations. We describe the detection of kink oscillations in Sect.~3. In Sect.~4, we perform the MHD seismology and estimate the localized physical conditions within the filament.  In Sect.~5, the discussions and conclusions are outlined.

\section{Observations}
\begin{figure}[htbp]

\centering
\includegraphics[width=1.\textwidth]{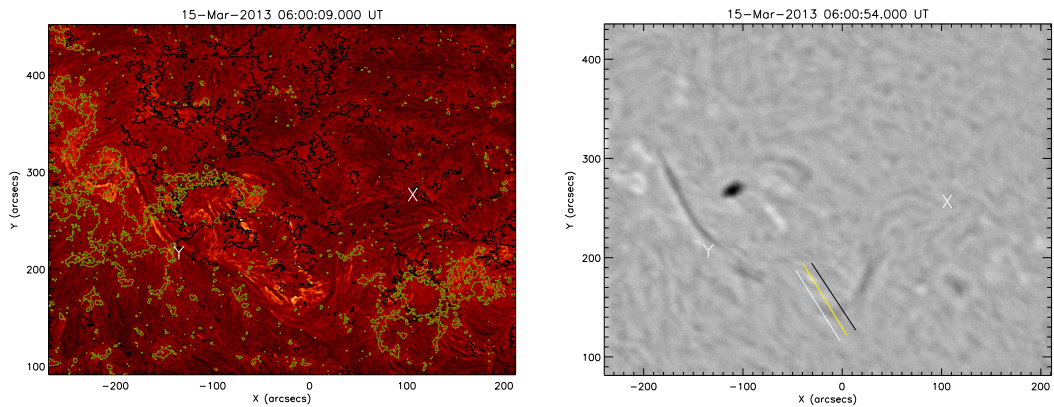}

\caption{The left panel shows the Helioseismic and Magnetic Imager (HMI) magnetogram contours overlayed on the image as observed by Atmospheric Imaging Assembly (AIA) on board SDO at 304~\AA~  channel. The right panel shows the {\it NSO}/{\it GONG} H$\alpha$ image of the filament  south-west to AR 11692. Slices used for creating time-distance map are marked as  white, yellow and black vertical lines respectively.}
\label{sun}
\end{figure}

An M1.1-class flare is observed by {\it GOES} satellite in the active region AR11692 on 2013 March 15, which is associated with a halo coronal mass ejection (CME). Soon after the onset of the flare, oscillations are observed in the filament lying in south-west to the active region as seen in H$\alpha$ from {\it NSO}/{\it GONG} (Fig.~\ref{sun}, bottom panel, the movie is available online.\\{\it NSO}/{\it GONG} provides full-disk observations of the Sun at 6563 \AA. It has a maximum pixel resolution of $\sim$1.07\arcsec~ and a cadence of 1 minute. The flare or the associated CME generated global disturbances that in turn triggered the observed transverse oscillations in the body of the filament. Observations started at 06:00:54 UT and ended at 10:22:54 UT. At least four transversal cycles of the global transverse oscillations are observed with their significant damping. The top panel of Fig.~\ref{sun}  shows the Helioseismic and Magnetic Imager (HMI) magnetogram contours overlayed on the image observed by Atmospheric Imaging Assembly (AIA) on board {\it SDO} at 304 \AA. AIA observes the Sun in seven Extreme Ultraviolet (EUV) bandpasses and has a pixel size of 0.6\arcsec \citep{lemen12}. The 304 \AA\ images are chosen because in this passband the footpoints of the filament can be identified. Footpoints are identified by examining the strands of the filament in the image together with contours of opposite polarities because a filament always lies on the polarity inversion line. Footpoints are marked as `X'  (negative) and `Y'  (positive) on the image.  Green and black contours represent constant magnetic field strength of 20 G and -20 G, respectively. We chose five points over the filament between the foot points `X' \& `Y'. We then interpolate the curve using cubic spline between the footpoints. The interpolated curve has to pass through the chosen five points. The length of the interpolated curve should be approximately equal to the length of the filament. We repeat this process several times by choosing different five points every time between `X' and `Y'. Therefore, the length of the interpolated curve will vary, thus its mean value and the standard deviation are estimated. The length of the filament, $L$, is measured to be $\sim235\pm$8 Mm.

\section{Detection of global kink oscillations in a filament tube}
\begin{figure}[htbp]
\centerline{\includegraphics[width=1.\textwidth,clip=]{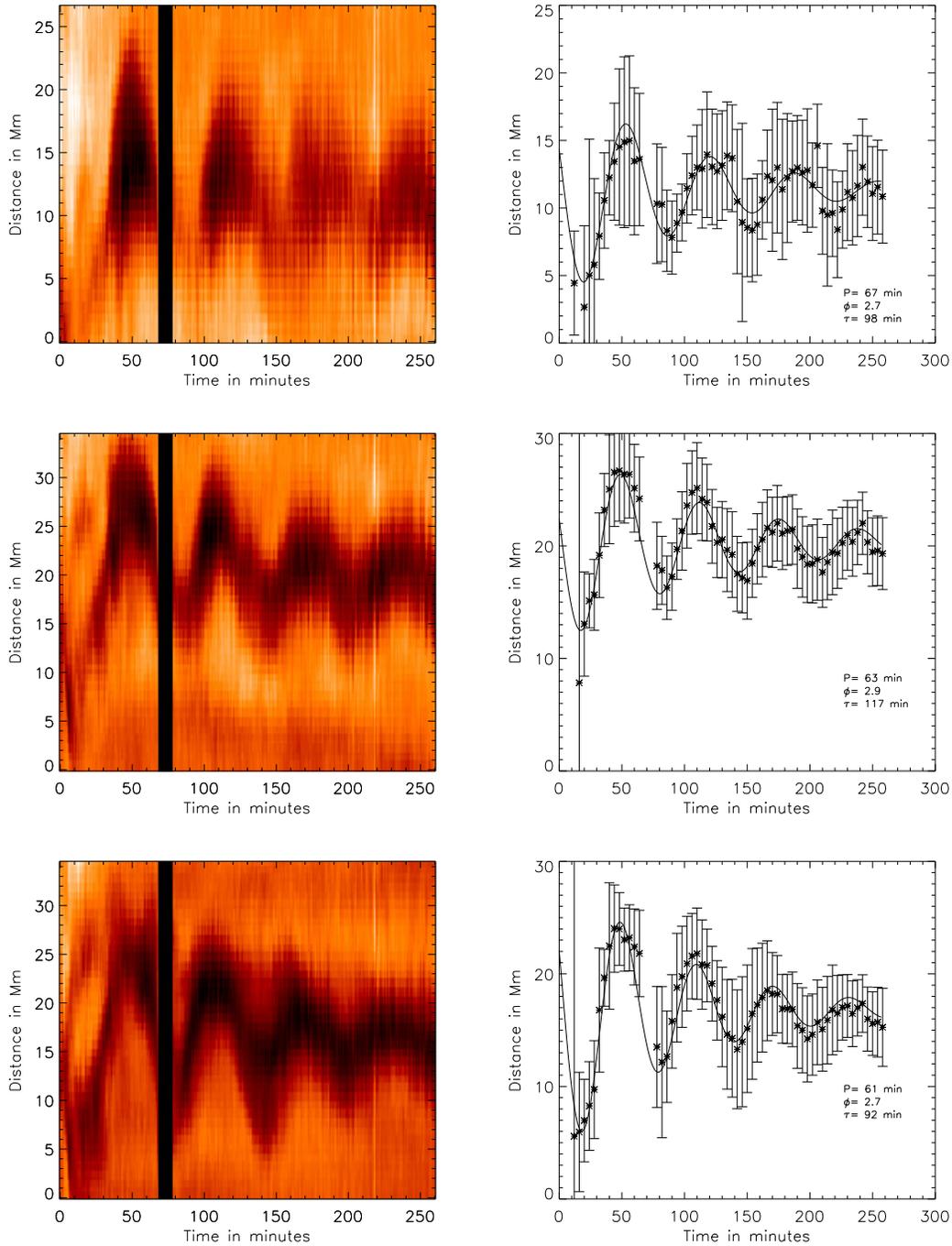}}
\caption{{\it Left}: time-distance maps (top to bottom) corresponding to the three slice locations as marked in Fig.~1 by white, yellow and black lines  respectively. {\it Right}: Exponentially damped sinusoidal curves}
\label{xt}
\end{figure}

The evolution of the oscillations is obtained by placing three artificial slices close to the apex of the filament parallel to the observed oscillations and perpendicular to the filament (see the right panel of Fig.~\ref{sun}) using the {\it NSO}/{\it GONG} H$\alpha$ image sequence. For each slice a two dimensional time-distance diagram is created (see the left panel of Fig.~\ref{xt}), where the $x$-axis represents the time in minutes and the $y$-axis represents the distance along the slice in Mm. Thick black region represents the data gap from 07:07:54 UT to 07:18:54 UT. Along each column of the time-distance diagram a Gaussian curve is fitted and the mean values with one-sigma error bars are estimated. The entire time-distance diagram is then fitted with a damped harmonic curve represented by the formula \citep{asch99} 
\begin{equation}
y(t)=c+a_{0}sin(\omega t+\phi)e^{-t/\tau_{d}},
\end{equation}
where $c$ is a constant, $a_{0}$ is the amplitude, $\tau_{d}$ is the damping time, $\omega$ is the angular frequency, and $\phi$ is the initial phase.

The least square fitting is performed using the function MPFIT.pro \citep{mark09} in the Interactive Data Language (IDL). The best-fit curve is shown in Fig.~\ref{xt} (right panel). The period of the oscillation $P=\frac{2\pi}{\omega}$, the damping time $\tau_{d}$, and the initial phase $\phi$ for the first (second and third) slices are found to be 67$\pm$3 min (63$\pm$2 min and 61$\pm$2 min), 98$\pm$47 min (117$\pm$45 min and 92$\pm$28 min) and 2.7$\pm$0.4 (2.9$\pm$0.3 and 2.7$\pm$0.3), respectively. From Fig.~\ref{xt} (right panel) we note that the filament segments at the three slices are oscillating in phase. The amplitude of the oscillation, $a_{0}$, is 8$\pm$4 Mm (9$\pm$3 Mm and 13$\pm$3 Mm) and the velocity amplitude is about 12$\pm$6 km s$^{-1}$ (15$\pm$6 km s$^{-1}$ and 22$\pm$7 km s$^{-1}$) for the first (second and third) slice, respectively. Oscillations with decaying amplitudes are also noticed near the footpoints of the filament. They are in phase with the oscillations seen near the apex of the filament. The in-phase displacement of the filament is interpreted as the global standing kink mode. Since all parts of the filament oscillate in phase and the footpoints of the filament are fixed, thus we infer it to be the fundamental standing mode. Moreover, it should be noted that we can not distinguish on the observational basis whether the oscillations are vertical or horizontal, or mixed. However, this does not affect the principle of the MHD seismology of the filament that we intend to implement in the present paper. Assuming the filament to be a flux tube embedded in uniform magnetic field, the phase speed for the fundamental standing mode can be calculated using the relation,
\begin{equation}
v_{ph}=\frac{2L}{P}.
\end{equation} 
As a result, $v_{ph}$ for the first (second and third) slice is computed to be 117$\pm$9 km s$^{-1}$ (124$\pm$8 km s$^{-1}$ and 128$\pm$9 km s$^{-1}$). The sound speed, $c_{s}$, with the chromospheric temperature is 15 km s$^{-1}$. With $v_{ph}>c_{s}$, we infer it to be the fast kink oscillation.

Prominence models in general invoke the fine structure of isolated/groups of prominence threads on which the MHD wave mode(s) distribute \citep{arre11}. In the present case, almost bipolar prominence having the aspect ratio rather small, like the EUV loops, is subjected to the transverse displacement of its axis as a whole. Therefore, we treat it as a loop-like prominence tube supporting the kink oscillations.

\section{Coronal Seismology of the Oscillating Filament Tube}
We assume the filament to be a flux tube with its radius much smaller than its length, $L$ (i.e., thin-tube approximation). With the thin tube approximation or long wavelength limit (i.e., $k_{z}a<<1$, where $k_{z}=2\frac{\pi}{\lambda}$ and $a$ is the cross-section of the filament, for the fundamental mode, $\lambda$=2$L$), the kink wave speed ($v_{k}$) is approximately equal to the phase speed ($v_{ph}$).

Analytical relation between the period of oscillation $P$, the internal Alfv\'en travel time $\tau_{A,i}$ and the density contrast $\zeta$ is given by
\begin{equation}
y=\frac{\tau_{A,i}}{P}=\frac{1}{\sqrt2}(\frac{\zeta}{\zeta+1})^{1/2},
\end{equation}
where $\tau_{A,i}=L/v_{A,i}$, the internal Alfv\'en speed $v_{A,i}=\frac{B}{\sqrt{\mu\rho_i}}$, $\mu$ is the permeability of the medium, and $\rho_{i}$ is the density inside the filament \citep{2008A&A...484..851G}. We take $\mu=4\pi$ X $10^{-7}$ H$m^{-1}$ in SI units. $\zeta$ in Eq.~(3)  is the internal to external density ratio $\frac{\rho_{i}}{\rho_{e}}$. From Eq.~(3), the Alfv\'en speed inside the filament tube is given by
\begin{equation}
v_{A,i}=\sqrt2\frac{L}{P}(\frac{\zeta+1}{\zeta})^{1/2}.
\end{equation}

The damping time of the fast kink waves may give an insight of ambient plasma and structure of the filament. \citet{arre08} proposed resonant absorption as a damping mechanism in the context of prominence fine structure oscillations. With the thin boundary (TB) approximation, the analytical expression of the damping time in the asymptotic limit in the Cartesian coordinates was derived by \citet{1988JGR....93.5423H} and later by \citet{1992SoPh..138..233G} and \citet{Ruderman2002}. The asymptotic expression is \citep[see][]{2006RSPTA.364..433G,arre08}
\begin{equation}
\frac{\tau_{d}}{P}=\frac{2}{\pi}\frac{R}{l}\frac{\rho_{i}+\rho_{e}}{\rho_{i}-\rho_{e}},
\end{equation}
where $R$ is the mean radius of the filament and $l$ is the thickness of its non-uniform layer. Sinusoidal variation of the density is assumed across its non-uniform layer leading to the factor of $\frac{2}{\pi}$ \citep{Ruderman2002} instead of $\frac{4}{\pi^2}$ for linear variations \citep{1992SoPh..138..233G}. Using Eq.~(5), we can express the inhomogeneity length scale ($z=l/2R$) as \citep{2008A&A...484..851G}
\begin{equation}
z=\frac{1}{C}\frac{\zeta+1}{\zeta-1},
\end{equation}
where $C$ is $\frac{\pi\tau_{d}}{P}$.

However, the first paper that consistently used  information on period and damping time is by \citet{arre07}. It used the full numerical results of the eigenvalue computations and its scheme is fully numerical.
From Eqs.~(4) and (6), we note that there are three unknown parameters $v_{A,i}$, $\zeta$, and $z$. There are infinite choices of $v_{A,i}$, $\zeta$, and $z$ available to satisfy Eqs.~(4) and (6). Therefore, we have to estimate one unknown and must express the other two unknowns in terms of it. We consider to fix the density contrast ($\zeta$) and express the Alfv\'en speed ($v_{A,i}$) and inhomogeneity length scale ($z$) in terms of $\zeta$. 

Maximum value of $\zeta$ is infinity. Since $z$ is the decreasing function of $\zeta$, $z$ attains the lowest possible value for the largest possible value of $\zeta$. Thus $z_{min}=\frac{1}{C}$ for $\zeta \rightarrow \infty$. The possible maximum value of $z$ is 1, therefore, the minimum value of $\zeta$ will be when $z$=1, which can be computed using Eq.~(6) as $\zeta_{min}=\frac{C+1}{C-1}$ and using Eq.~(4) as $v_{A,i max}=\frac{2L}{P}\sqrt{\frac{C}{C+1}}$. In the next step, we compute the magnetic field strength and inhomogeneity length scale separately corresponding to slices I, II and III for a fixed  value of $\zeta$.

\subsection{Density estimates using automated DEM analysis}
It is evident that the Alf\'ven speed in a medium depends on magnetic field strength and density of the medium. Using Eq. (4),  Alf\'ven speed can be estimated for a given value of $\zeta$. Therefore, magnetic field inside the filament can be estimated if density of the filament is known. To estimate density values we follow automated temperature and emission measure analysis technique as developed by \citet{asch2013} to derive the average density and temperature inside the filament tube. Using this technique the electron density ($n_{e}$) and temperature ($T_{e}$) inside the filament is estimated to be $10^{8.60\pm 0.32}$ cm$^{-3}$ and $10^{5.93\pm 0.25}$ K (see Fig.~3). Note that the density is underestimated as  compared to the typical density range of prominences which ranges from $10^{9}$ to $10^{11}$ cm$^{-3}$ for quiescent and active filaments \citep{2010SSRv..151..243L}. We feel that  because of the low density contrast with respect to the background as visible in various {\it SDO}/AIA channels, the DEM method does not work very well. Furthermore, the observed filament is made-up of mostly cooler material, whereas various AIA channels are sensitive to hotter plasmas, thus the contribution from the prominence material would be minimal. The bright points seen in AIA 171 \AA~ channel (see Fig.~3) correspond to the prominence-corona transition region (PCTR). Thus the density values calculated by the DEM technique is the density of the PCTR associated with this prominence, rather than the bulk of the prominence material. Thus the DEM forward fitting will not be accurate enough for the prominence. By DEM analysis we wish to estimate the PCTR magnetic field. If  we use these density estimates (from DEM) and assuming the number density ratio H:He to be 10:1 inside the filament, the magnetic field strength ($B$) for density contrast $\zeta$ =100 are estimated as 0.86$\pm 0.21$ G (0.91$\pm 0.20$ G and 0.94$\pm 0.20$ G) for first (second and third) slice respectively. These magnetic field strengths are much lower than as reported in \citet{2010SSRv..151..333M} for the quiescent filaments (3--15 G). It implies that the magnetic field of overlying PCTR is lesser compared to typical magnetic field of filament.
\begin{figure}[htbp]
\centerline{\includegraphics[width=1.\textwidth]{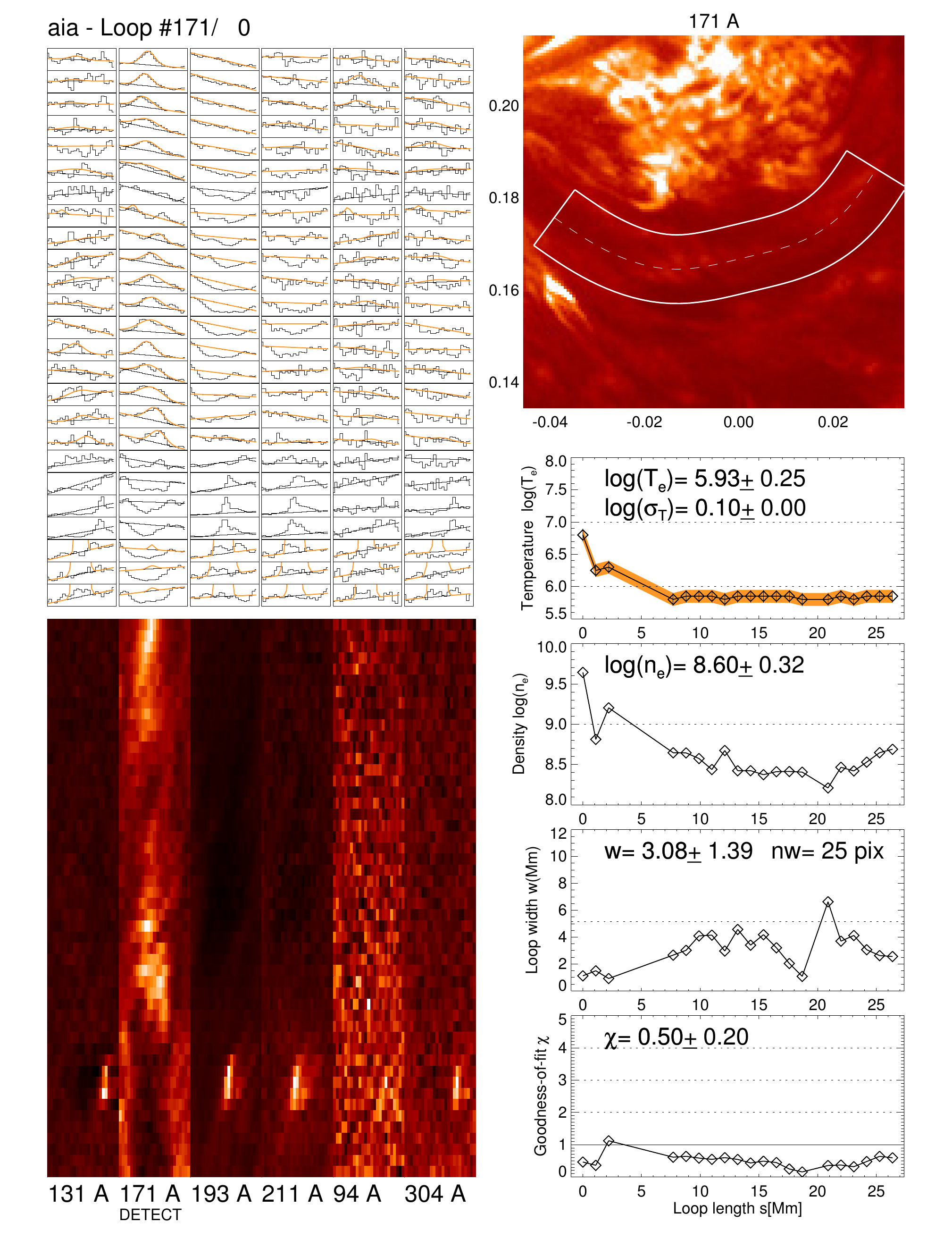}}
\caption{Automated DEM analyses measurements : (i) the selected segment of the tube (top-right)
which is also shown in various channels of the AIA (bottom-left). (ii) The forward DEM fitting
over the emission derived along the chosen path (top-left), and derived density as well as 
temperature (blocks in bottom-right multipanel plot).
 }
\label{dem}
\end{figure}

\subsection{Magnetic field and inhomogeneity length estimation corresponding to Slice I}
Since automated DEM analysis underestimate the density values thus we take typical density values in the filament channel which ranges from $10^{9}$ to $10^{11}$ cm$^{-3}$. Using the observed period $P$=67$\pm$2 min as well as the damping time $\tau_{d}$=98$\pm$47 min, we get $\frac{\tau_{d}}{P}$=1.46$\pm$0.76. For simplicity, henceforth in the paper we consider the absolute values of the parameters without error bars. For example, we derive the extreme values of the seismic quantities as  $C$=4.59, $\zeta_{min}$=1.557, $z_{min}$=0.218 and $v_{A,i,max}$=106 km s$^{-1}$. Table ~\ref{T-complex} shows the value of $v_{A,i}$ and $z$ for different values of $\zeta$. There are infinite choices as shown in Table~1, but we take typical value of $\zeta$ for the filament to be 100. For $\zeta$=100 the Alfv\'en velocity and $z=\frac{l}{2R}$ are 83.08 km s$^{-1}$ and 0.222, respectively, thus $l/R$=0.444. For typical density values in the filament channel ($10^{9}$ to $10^{11}$ cm$^{-3}$), we deduce the magnetic field value to be $\sim$1.36 G, $\sim$4.28 G, and $\sim$13.6 G, respectively.

 \begin{table}
\begin{center}
\caption{Seismic parameters for exponential damping} 

\label{T-complex}
\begin{tabular}{lcccc r@{   }l c} 
  \hline
  & \multicolumn{2}{c}{Slice I} & \multicolumn{2}{c}{Slice II} & \multicolumn{2}{c}{Slice III} \\
 $\zeta$ & $v_{A,i}$ & $z$ &      $v_{A,i}$ & $z$ &            $v_{A,i}$ &  $z$\\
  \hline
\hline
1.414&-- &-- & 115&1&--&--\\
1.534&--&--&113&0.814&116.69&1\\
1.557 & 106 & 1 &112.67 &0.787 &116.37 &0.968 \\
2 & 101.25 & 0.654 & 107.68&0.515&111.21&0.633\\
3&95.46&0.436&101.52&0.343&104.85&0.422\\
4&92.42&0.363&98.30&0.286&101.52&0.352\\
5&90.56&0.327&96.31&0.257&99.47&0.316\\
6&89.29&0.305&94.97&0.240&98.07&0.295\\
7&88.38&0.290&94&0.229&97.07&0.281\\
8&87.69&0.280&93.25&0.220&96.31&0.271\\
9&87.14&0.272&92.68&0.214&95.72&0.264\\
10&86.71&0.266&92.21&0.210&95.24&0.258\\
50&83.49&0.227&88.79&0.179&91.71&0.220\\
100&83.08&0.222&88.36&0.175&91.26&0.215\\
1000&82.71&0.218&87.97&0.172&90.85&0.211\\
.&.&.&.&.&.&.\\
.&.&.&.&.&.&.\\
$\infty$&82.67&0.218&87.92&0.171&90.80&0.211

\end{tabular}
\end{center}
\end{table}

\subsection{Magnetic field and inhomogeneity length estimation corresponding to Slice II}
Using the observed period $P$=63$\pm$2 min and damping time-scale $\tau_{d}$=117$\pm$45 min, we get $\frac{\tau_{d}}{P}$=1.86$\pm$0.77. The seismic quantities $C$, $\zeta_{min}$, $z_{min}$ and $v_{A,i,max}$ are calculated as 5.83, 1.414, 0.171, and 115 km s$^{-1}$, respectively. For $\zeta$=100, the Alfv\'en velocity and $z=\frac{l}{2R}$ are 88.36 km s$^{-1}$ and 0.175, respectively, thus $l/R$=0.350. Assuming similar densities values as those for slice I, i.e., $n_{e}$= $10^{9}$, $10^{10}$ and $10^{11}$ cm$^{-3}$, the magnetic field strength $B$ for the given density contrast $\zeta$ =100 is estimated to be $\sim$1.44 G, $\sim$4.56 G, and $\sim$14.46 G, respectively. 

\subsection{Magnetic field and inhomogeneity length estimation corresponding to Slice III}
Using the observed period $P$=61$\pm$2 min and damping time $\tau_{d}$=92$\pm$28 min, we get $\frac{\tau_{d}}{P}$=1.51$\pm$0.51. The seismic quantities $C$, $\zeta_{min}$, $z_{min}$, and $v_{A,i,max}$ are calculated as 4.74, 1.534, 0.211, and 116.69 km s$^{-1}$, respectively. For $\zeta$=100 the Alfv\'{e}n velocity and $z=\frac{l}{2R}$ are 91.26 km s$^{-1}$ and 0.215, respectively, thus $l/R$=0.430. Assuming similar densities as those for slice I, i.e., $n_{e}$= $10^{9}$, $10^{10}$ and $10^{11}$ cm$^{-3}$, the magnetic field strength $B$ for the given density contrast $\zeta$ =100 is estimated to be $\sim$1.48 G, $\sim$ 4.70 G, and $\sim$ 14.85 G, respectively. 

\subsection{Magnetic field and inhomogeneity length estimation for infinite density contrast}

Seismology is performed by using both P and $\tau_D$ and the two coupled equations (Eqs. 4-6). We keep the one unknown, i.e, density contrast fixed, and express the Alfv\'en speed and inhomogeneity length scale in terms of it. In the case of prominence plasmas, if we consider  the density contrast is large and tending towards infinity, we get two uncoupled equations in which one relates the period and Alfv\'{e}n speed, while other the damping ratio with the inhomogeneity length scale ($l/R$).  Therefore, the inversion what we performed earlier is independent of the density contrast. This is clearly evident from Table 1 which shows that  various  parameters are having almost similar values for higher values of $\zeta$ (=1000, $\infty$). For $\zeta$=$\infty$ for slice I, II, III, the Alfv\'en velocities and $z=\frac{l}{2R}$ are (82.67, 87.92, 90.8) km s$^{-1}$ and (0.218, 0.171, 0.211), respectively, thus $l/R$=(0.436, 0.342, 0.422). The magnetic field strength estimated for $\zeta$=$\infty$ is similar to what is estimated by assuming $\zeta$=100 to first place of decimal.


\section{Summary and Discussion}\label{SECT:DISS}
A dynamical study of the cool filament observed on-disk in H$\alpha$ has been performed. We report global kink oscillations in the filament system as observed in H$\alpha$ as the whole filament tube oscillates in the transverse mode. The oscillation is most likely excited by the large-scale disturbances originating from a nearby M1.1-class flare. However, we do not aim to understand about the nature of the driver and its properties, it is out of the scope of our present article.
Time series analysis revealed a period between 61 to 67 min and a damping time between 92 to 117 min associated with the kink oscillations around the apex of the filament system. The phase speed ($v_{ph}$) is found to be between $\sim$117 km s$^{-1}$ to 128 km s$^{-1}$. These are long-period damped fast kink oscillations, as the phase speed is high compared to the chromospheric sound speed.
We also find that oscillations are damped. \citet{Lin2009} has carried out prominence seismology in prominence threads and estimated the magnetic field of about 0.1--20 G. However, We have carried out consistent seismology using the values of the periods and the damping times at the same time. It is worthwhile to emphasize the use of seismology using the values of the periods and the damping times as it gives an order of estimate of upper limit of Alfv\'en speed inside the prominence \citep{2008A&A...484..851G} and hence on the estimation of the magnetic field strength. \citet{Pinter2008} carried out the prominence seismology and calculated the axial component of the magnetic field in a polar crown prominence to be 1--5 G. They used twisted flux rope model to calculate the magnetic field values. To the best of our knowledge, there have not been any reports yet on consistent seismology in large amplitude prominence oscillations, which we implement in the present work to diagnose the local plasma conditions of the observed filament tube. 
The main result from this study is the estimation of the inhomogeneity length scale. Assuming a typical density ratio $\zeta$ of filament to be 100, the inhomogeneity length scale, $l/R$ is 0.444 (0.350 and 0.430) for the first (second and third) slice, respectively, which suggests that the inhomogeneous layer is quite thick near the apex of the filament. The density ratio estimate is not very accurate, since there are infinite choices, so from the available information we can say that $l/R$$>$ 0.436 (0.342 and 0.422) for the first (second and third) slice, respectively. We find the magnetic field strength, using the typical density values, in the filament to be in the range of $\sim$1 G to 15 G, which is close to what has been reported by \citet{Lin2009} for the prominence threads. To best of our knowledge, this is the first time we estimated inhomogeneity length scale and magnetic field strength of a filament as a whole using period of oscillation and damping time simultaneously.
\citet{Ruderman2013} reported that the amplitude decay in standing kink oscillations for sufficiently small time is described by a Gaussian function and becomes exponential for a later time. The transition from Gaussian to exponential depends on the inhomogeneity length scale. Similar studies have been carried out by \citet{Pascoe2012} and \citet{Hood2013} for propagating waves. Since the knowledge of inhomogeneity length scale is not known {\it a priori} thus we use Gaussian profile to investigate how it affects the seismology parameters. We find that Alf\'ven speed and magnetic field strength are almost same as calculated using exponential damping. However, inhomogeneity length scale is different because the Gaussian function falls slowly than exponential function and therefore damping time is large. Therefore, we conjecture that based on our MHD seismology that inhomogeneous layer was already present when the global kink oscillations are excited. This further causes the quick damping of the oscillations via the resonant absorption through the well evolved layer of inhomogeneity across the prominence tube.

\section{Acknowledgments}
We would like to thank the Referee for his/her valuable comments which has enabled us to improve the manuscript. M.G. acknowledges support from KU Leuven via GOA/2009-009 and also partial support from the Interuniversity AttractionpPoles Programme 
initiated by the Belgian Science Policy Office (IAP P7/08 Charm). PFC is supported by the Chinese foundations 2011CB811402 and NSFC (11025314,
10933003, and 10673004).


{}

\clearpage

\end{document}